\documentclass{nature}
\usepackage{ifthen}
\usepackage{graphicx}
\usepackage{amsmath}
\usepackage{amssymb}
\usepackage{color}

\newboolean{SubmittedVersion}
\setboolean{SubmittedVersion}{False}
\DeclareMathAlphabet{\mathpzc}{OT1}{pzc}{m}{it}

\def\Rb87{^{87}\text{Rb}}
\def\Na23{^{23}\text{Na}}
\def\Li6{^{6}\text{Li}}

{\catcode`\|=\active
  \gdef\Braket#1{\left<\mathcode`\|"8000\let|\BraVert {#1}\right>}}
\def\BraVert{\egroup\,\mid@vertical\,\bgroup}

\begin{document}

\title{Stability of similarity measurements for bipartite networks}
\author{Jian-Guo~Liu$^{1,\ast}$, Lei~Hou$^1$, Xue~Pan$^1$, Qiang~Guo$^1$, Tao~Zhou$^{2,\ast}$}
\maketitle

\begin{affiliations}
\item
1, Research Center of Complex Systems Science, University of Shanghai for Science and Technology, Shanghai 200093, PR China. \\
2, CompleX Lab, Web Sciences Center and Big Data Research Center, University of Electronic Science and Technology of China, Chengdu 611731, PR China. \\
$^\ast$ To whom correspondence should be addressed. E-mail: liujg004@ustc.edu.cn, zhutou@ustc.edu
\end{affiliations}

\begin{abstract}
Similarity measurements are widely used in the user-object bipartite networks to evaluate potential relations between objects. We argue that the accurate similarity measurements should generate stable results for objects since their natural properties are unchangeable regardless of the network structure. With six bipartite networks, the present paper quantifies the stabilities of fifteen similarity measurements by comparing the similarity matrixes of two data samples which are randomly divided from original data. Results show that, the fifteen measurements can be well classified into three clusters according to their stabilities. Measurements in the same cluster are found having the same considerations and similar mathematical definitions. In addition, we develop a top-$n$-stability method to study the object similarity stability's effect on the recommendation. The unstable similarities are proved to be false information By taking only the stable similarities into account, the stability of the recommendation could be largely improved. Our work may shed some lights on the further investigation and application of similarity in link prediction and recommendation systems.
\end{abstract}

Connections are everywhere and can be observed between everything in our world \cite{LINKED}, such as the connections between locations in the transport systems \cite{TRANSPORT1, TRANSPORT2} and connections between neurons in the neural networks \cite{NEURAL}. To characterize systems consisting of connections, complex network, in which objects are simplified as nodes and connections are simplified as links, has been widely used in the last decade to study the relations between different objects and the structure of those kinds of systems \cite{small-world, BA, review-network}.

However, in most real complex networks, not only the connected nodes have relations with each other. Actually, every pair of nodes has some specific relations even there is no link between them. For example, citation networks consist of papers and if a paper cited another, there would be a link between them \cite{CITATION1}. Besides citing relations between connected papers, other potential relations may exist between any pair of papers like same subject, same author or same cited papers and so on \cite{CITATION2}. Generally, {\it similarity} that describes the connections between different objects' properties, is the most used method to evaluate such relations. With the rapid development of complex networks, similarity has become an important measurement with great significance for both theoretical research in fields such as biological and physical science and practical applications in e-commerce and social service. Evaluating the similarity of genes' expression profile, one may identify similar regulations and the control processes of genes \cite{GENE1, GENE2}. Co-expression networks may also be established according to the similarities between genes \cite{GENE3, GENE4}. Moreover, while protein-protein or metabolic interactions can only be verified by costly experiments and most of the interactions are still unknown, similarity-based link prediction method \cite{LINK_PRE1, LINK_PRE2} could largely help identify the most likely pair of interacting proteins \cite{PROTEIN1, PROTEIN2, PROTEIN3}. In addition, the similarity measurements have also found its' applications in object clustering \cite{CLUSTERING1, CLUSTERING2} and community detecting \cite{COMMUNITY}. As to more practical applications, one of the most used recommendation systems \cite{RECOMMENDATION1, RECOMMENDATION2, RECOMMENDATION3} is to utilize similarity to evaluate correlations between objects such as movies, commodities, books and so on, and accordingly make recommendations to users.

Based on various theories and considerations, dozens of similarity measurements have been developed. However, with different data, similarity measurements generally have very different performances. Even with different parts of a same data, the results may also vary. Particularly in the bipartite networks, the object similarity is determined by their natural properties, and thus, similarity should be steadfast for a definite pair of objects. On the other hand, the networks we study are mapped incompletely, which is always evolving, or contain false positives and negatives \cite{RANKING_STA}. Thus, some fundamental questions are roused that, how dose the network structure affects the similarity? Could those exists similarity indexes describe the real relations between objects? While the aim of similarity measurements is to estimate the real correlation between items, unstable estimation is unreliable and meaningless. For a definite pair of objects, a good measurement should always return the same similarity at different times. To explore the stability problem of object similarity in bipartite networks, fifteen similarity measurements will be analyzed and studied in this paper. Firstly, we will report the influence of data amount on the fifteen similarity measurements' stability. Secondly, the comparison and classification of the fifteen similarity measurements will be analyzed. At last, we will explore the object similarity stability's effect on the recommendation.

\begin{table*}[!htb]
{
\caption{Properties of the utilized datasets. The sparsity is the deviation between existed links and possible links, i.e. $T/(M\cdot N)$, where $T$, $M$, $N$ is the number of links, objects and users respectively. Subject matters of those datasets are definite objects whose properties are unchangeable except Last.FM. The subjects of Last.FM are artists. However, the artists' music have definite contains and properties. Thus, the artists in Last.FM could also be regarded as objects.}
\begin{center}
\begin{tabular}{llrrrr}
   \hline\hline
    Dataset & Subject matter & NUM. of users & NUM. of objects  & NUM. of links & Sparsity \\
   \hline
    {\it MovieLens}   & Movie     & 5547   & 5850  & 698054  & $2.15\times 10^{-2}$\\
    {\it Netflix}     & Movie     & 8608   & 5081  & 419247  & $9.59\times 10^{-3}$\\
    {\it Amazon}      & Commodity & 645056 & 99622 & 2036091 & $3.17\times 10^{-5}$\\
    {\it Last.FM}     & Artist    & 1892   & 17632 & 92834   & $2.78\times 10^{-3}$\\
    {\it Epinions}    & Reviews   & 28090  & 30073 & 422085  & $5.00\times 10^{-4}$\\
    {\it Del.cioi.us} & Bookmark  & 1861   & 1860  & 15328   & $4.43\times 10^{-3}$\\
   \hline\hline
\end{tabular}
\end{center}
}
\end{table*}

\vspace{\baselineskip}
{\large\noindent\textbf{Similarity Measurements}}\\
In many online systems objects usually could get ratings from different users. To this kind of context, one can use the {\it Cosine Index} (COS) or {\it Pearson Coefficient} (PC) to measure the object similarity. When the ratings are unavailable, similarity can also be defined from the structure of the historical data, that is, two objects are similar if they are connected with many same nodes. The simplest such method is {\it Common Neighbor} (CN), where the similarity between two objects are directly given by the number of same neighbors who have connections with them. Considering degree information of two objects, variations of CN have been proposed, including {\it Salton Index} (SAL) \cite{SAL}, {\it Jaccard Index} (JAC) \cite{JAC}, {\it S{\o}rensen Index} (SOR) \cite{SOR}, {\it Hub Promoted Index} (HPI) \cite{HPI}, {\it Hub Depressed Index} (HDI) and {\it Leicht-Holme-Newman Index} (LHN) \cite{LHN}. Instead of number of the same neighbors in the CN index, {\it Adamic-Adar Index} (AA) \cite{AA} and {\it Resource Allocation Index} (RA) \cite{RA} was presented, regarding objects' similarity as the summation of their common neighbors' degree information. To fit the basic preferential attachment rule in network science \cite{BA}, the algorithm {\it Preferential Attachment Index} (PA) was also presented. Furthermore, using the concepts from physics, the method {\it Mass Diffusion} (MD) \cite{MD}, {\it Heat Conduction} (HC) \cite{HC1, HC2} and {\it Improved Heat Conduction} (IHC) \cite{IHC} were also investigated. The mathematical definitions of those similarity measurements can be found in the Method section. Generally speaking, the value of similarity is relatively high (low) if the objects are very similar (different). With these fifteen similarity measurements, we investigate their stability when measuring object similarity in the user-object bipartite networks.

\begin{figure*}[!htb]
\center\scalebox{.85}[.85]{\includegraphics{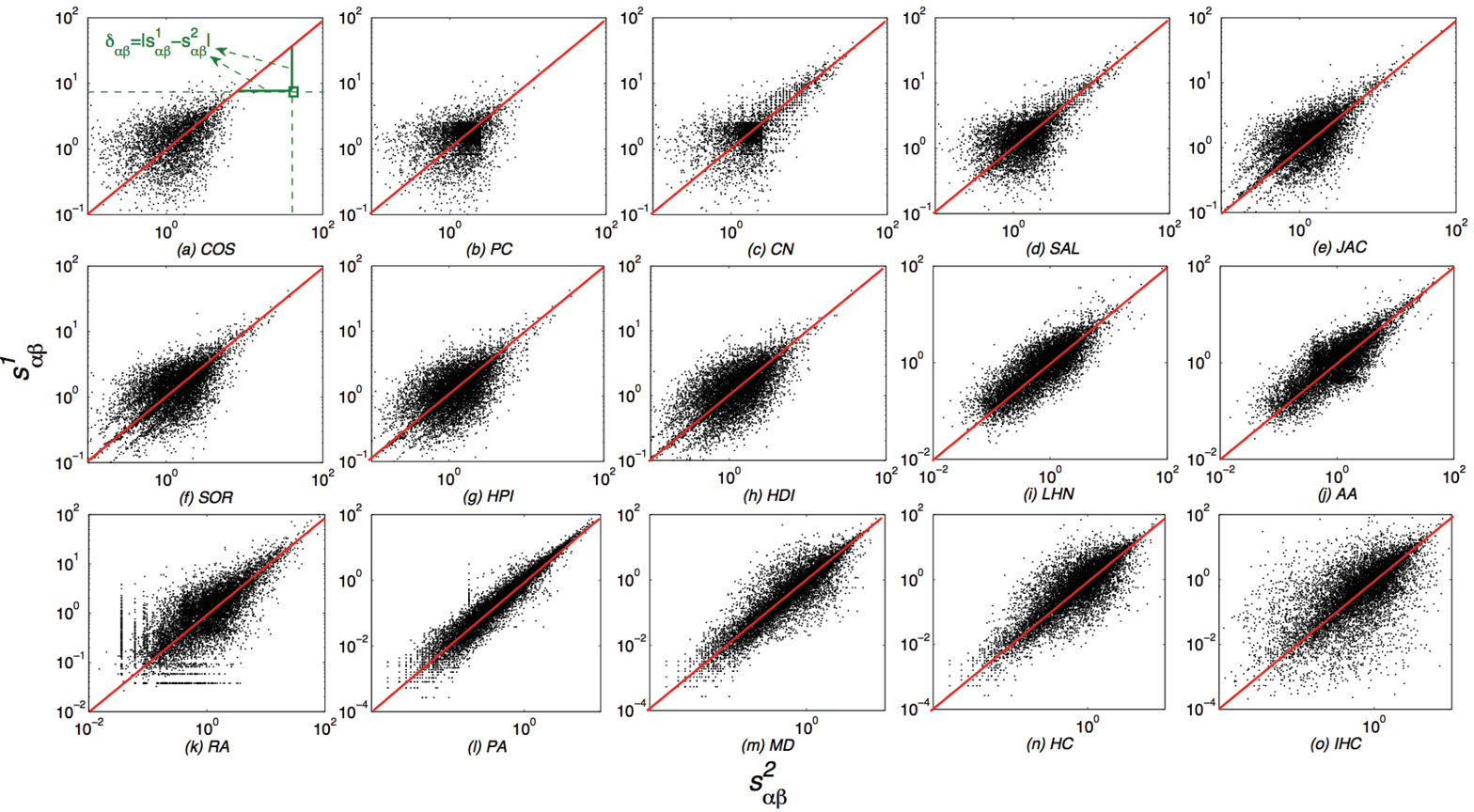}}
\caption{(Color online) Typical examples of the comparison between object similarities $s_{\alpha\beta}^1$ and $s_{\alpha\beta}^2$ of MovieLens dataset for fifteen similarity measurements. When randomly dividing data, we have $\eta=0.5$, i.e. 50\% ratings are divided into the first sample and the others are divided into the second sample. For each calculation, we randomly select $10^4$ pairs of objects' similarities of two sample to show in the figure. Thus, there are $10^4$ dots in each subplot, each representing a pair of objects. The dots would locate on the diagonal if the similarities in two samples $s_{\alpha\beta}^1=s_{\alpha\beta}^2$. Consequently, the more stable the similarity measurement is, the more concentrated the dots would distribute around the diagonal.}
\end{figure*}

\vspace{\baselineskip}
{\large\noindent\textbf{Data}}\\
Six different datasets are applied in this paper to study the stability of similarity measurements, differing both in the subject matter and data sparsity, as shown in Table 1. Those datasets are widely used to investigate and evaluate the recommendation algorithms and usually modeled as the user-object bipartite networks \cite{BIPARTITE1, BIPARTITE2, BIPARTITE3}. {\it MovieLens} and {\it Netflix} are movie Web sites in which users could watch and rate movies. {\it Amazon} is a e-commerce systems in which users buy commodities. {\it Last.FM} is a music Web site allowing users collect different artists' music. {\it Epinions} allows users writing reviews and on the other hand reading others' reviews. {\it Del.icio.us} is a bookmark Web site in which users collect and share bookmarks they interested in.

\vspace{\baselineskip}
{\large\noindent\textbf{Similarity Stability}}\\
Although lots of object similarity measurements have been presented, we could not know the exact object similarity. Thus, to examine the stability of those measurements, we divide the dataset into two samples to compare the similarity matrixes calculated from those two samples for each measurement. The data-dividing method can be described as follow: Every record will get a random number $p$ from a uniform distribution ranging from 0 to 1, and this record belongs to the first sample if $p \leq \eta$ and else belongs to the second sample if $p > 1-\eta$, where $\eta$ can be regarded as a data amount parameter and $0<\eta\leq0.5$. With this method, those two samples would have no overlaps, which means, they are totally different parts of the dataset. For a specific pair of objects $\alpha$ and $\beta$, we use $s_{\alpha\beta}^1$ to denote their similarity in the first sample and $s_{\alpha\beta}^2$ to denote that in the second sample. Thus, if a similarity measurement can give stable evaluation of the object similarity, there would be $s_{\alpha\beta}^1=s_{\alpha\beta}^2$. Figure 1 reports the distributions of similarities of two samples for each of the fifteen similarity measurements in MovieLens dataset. The dots would distribute near the diagonal if the measurement can give stable evaluation of object similarity. The PA index presents the most concentrated distribution. The reason is, the PA index only considers the popularity when measuring the objects. Popular objects of a data sample are in general also popular in another data sample, and thus the object similarity is stable. Other measurements' results are not so concentrated especially for pairs of objects with low-similarity pairs of objects. Results in Fig. 1 indicate that, when the data is changed, a same pair of objects may be evaluated as different similarity level and thus, the stability problem indeed exists in most similarity measurements.

\begin{figure*}[!htb]
\center\scalebox{1}[1]{\includegraphics{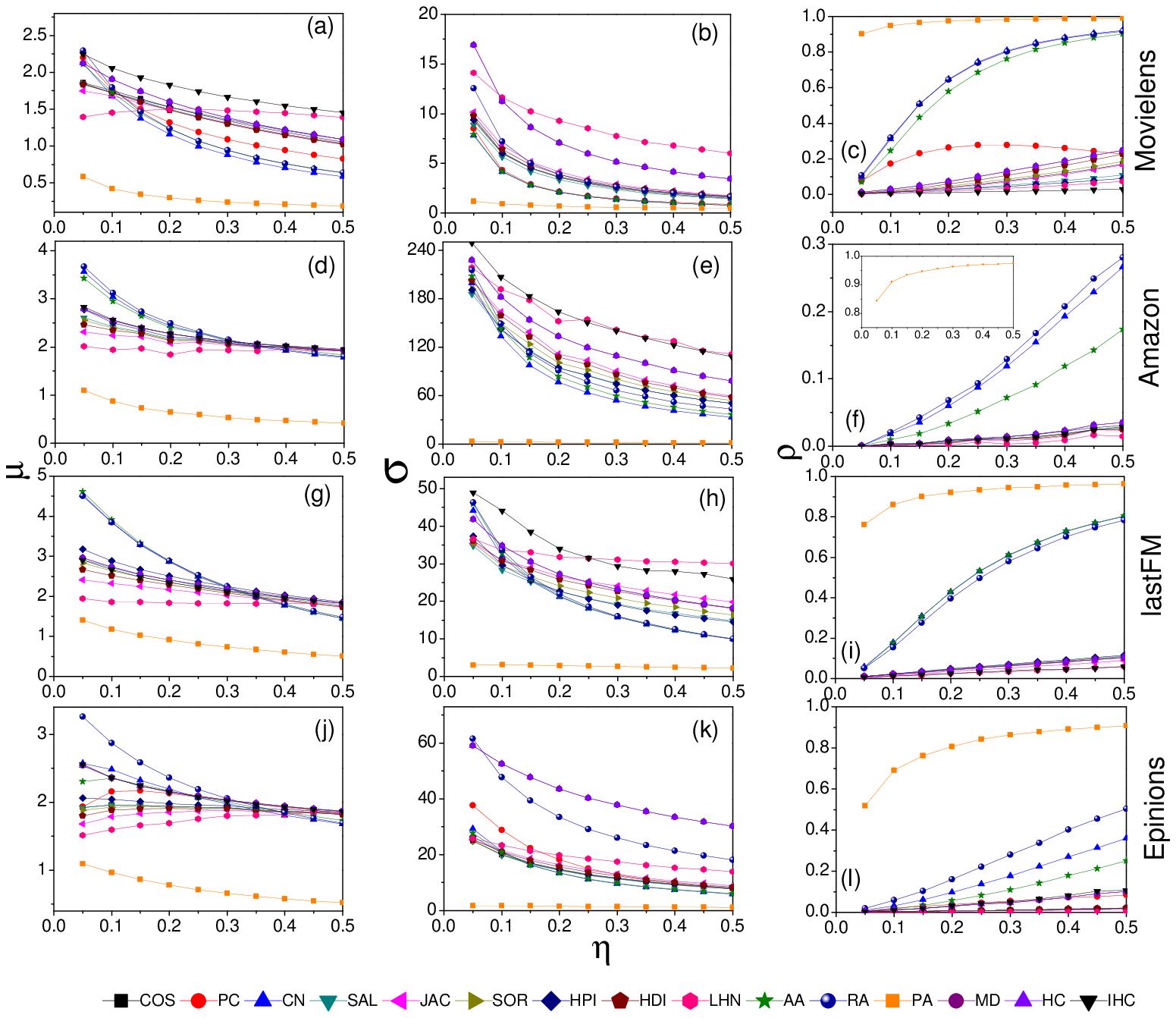}}
\caption{(Color online) Average bias $\mu$, standard deviation of bias $\sigma$ and the Pearson coefficient $\rho$ against the data amount parameter $\eta$ for MovieLens, Amazon, Last.FM and Epinions data respectively. Each data point is averaged over 20 independent experiments, i. e., for each level of data amount parameter $\eta$, we randomly divide the data for 20 times and calculate $\mu$, $\sigma$ and $\rho$ of each time. Note that, there is only selecting information without ratings in Last.FM dataset. Thus, the COS and PC indexes could not be performed in Last.FM dataset. As the data becomes more and more abundant, the stability of object similarity would relatively be better. However, many measurements still could not give stable evaluations of object similarity.}
\end{figure*}

\begin{figure*}[!htb]
\center\scalebox{0.8}[0.8]{\includegraphics{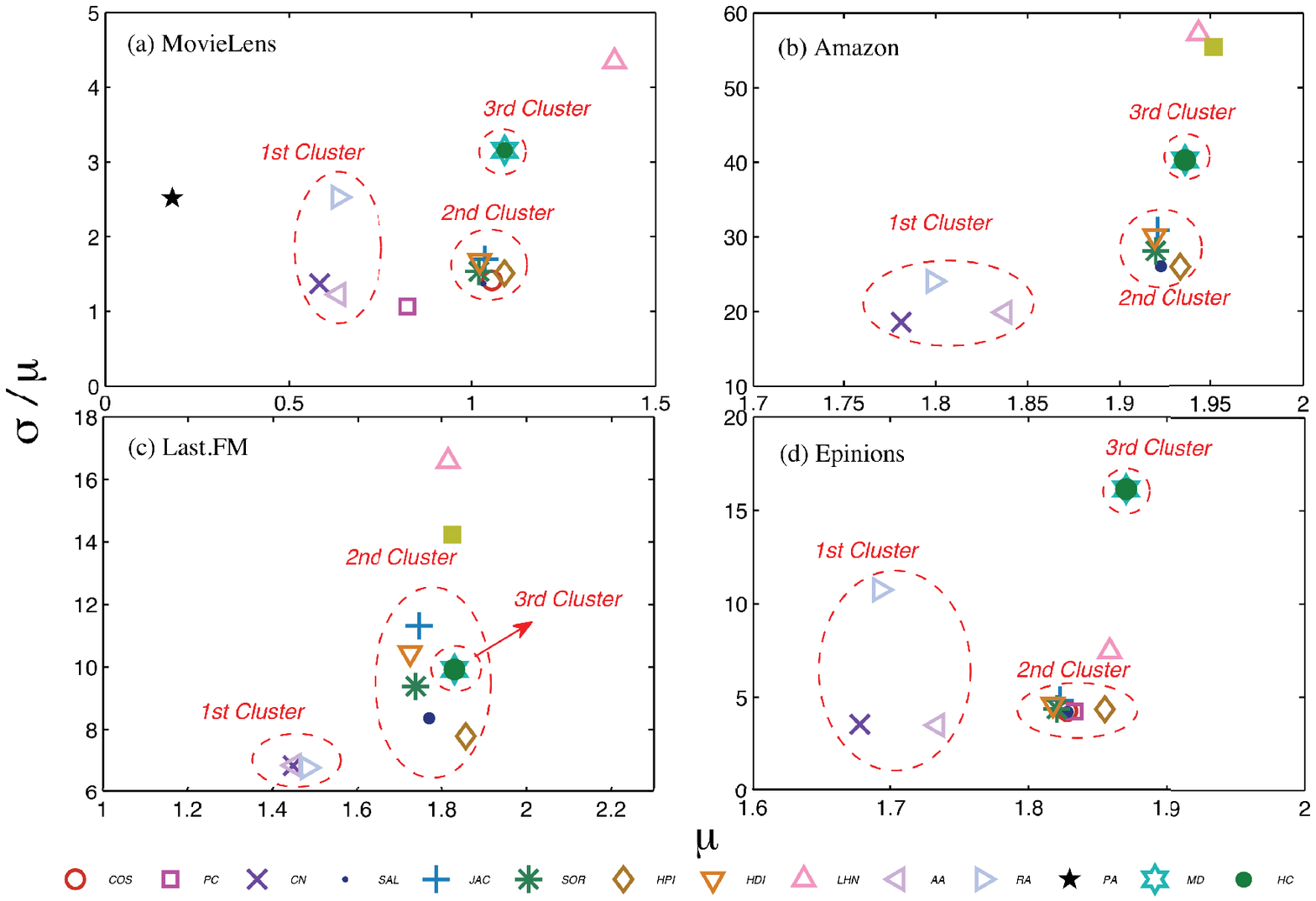}}
\caption{(Color online) The $\mu-\sigma$ location map with data amount parameter $\eta=0.5$. On the location map , a measurement locating on the left side means the similarities of objects have little change at average, and the bottom means the similarities of each pair objects have similar change. Overall, a stable measurement generally would locate on the left bottom of the $\mu-\sigma$ location map.}
\end{figure*}

The values of similarity calculated by different measurements distribute in different ranges, and thus we make a simple normalization to make those measurements comparable. Suppose that $\overline{s}$ is the average value of similarity that $\overline{s}=\sum_{\alpha\beta} s_{\alpha\beta}/(N(N-1))$ where $N$ is the number of objects which have at least one record in the corresponding sample, the normalization is $s_{\alpha\beta}=s_{\alpha\beta}/\overline{s}$. Specifically, the similarities of the PC index distribute in the range $[-1, 1]$, which may probably leads $\overline{s}$ to be 0. Hence, we make the normalization as $s_{\alpha\beta}=(s_{\alpha\beta}+1)/(\overline{s}+1)$ for the PC index. Henceforth, the similarities are all been normalized before used. To qualify the stability of object similarity, we define three metrics:

1) The average bias $\mu$ is used to describe the average level of similarity difference between two similarity matrixes from the two samples, and it reads
\begin{equation}
\mu=\frac{\sum_{\alpha\beta}\delta_{\alpha\beta}}{N(N-1)},
\end{equation}
where $\delta_{\alpha\beta}$ is the bias of similarities between objects $\alpha$ and $\beta$ from two samples as shown in Fig. 1 (a), i.e. $\delta_{\alpha\beta}=|s_{\alpha\beta}^1-s_{\alpha\beta}^2|$. High value of average bias means the same pair of objects is evaluated as different similarity level on average when the data is changed. Thus, the more stable the similarity measurement is, the lower the value $\mu$ would be.

2) The standard deviation of bias $\sigma$ reads
\begin{equation}
\sigma=\sqrt{\frac{\sum_{\alpha\beta}(\delta_{\alpha\beta}-\mu)^2}{N(N-1)}}.
\end{equation}
The deviation $\sigma$ can measure the difference of susceptibility between different pairs of objects against the data change. High values of the deviation $\sigma$ mean that, similarities between some pairs of objects may be quite unstable. On the other hand, low values of $\sigma$ could indicate that, each pair of objects has similar unstable level and the bias $\mu$ may be caused by the coincident entirety changes of every pair of objects' similarity.

3) The Pearson coefficient $\rho$ reads
\begin{equation}
\rho=\frac{\sum_{\alpha\beta}(s_{\alpha\beta}^1-\overline{s}^1)(s_{\alpha\beta}^2-\overline{s}^2)}{v_1v_2N(N-1)},
\end{equation}
where $\overline{s}^1$ and $\overline{s}^2$ are the average value of similarity over every pair of objects for two samples respectively, and $v_1$ and $v_2$ are the standard variance of similarity for two samples respectively. In general, the value of Pearson coefficient $\rho$ ranging from -1 to 1 measures the coherence of two similarity matrixes calculated by two samples. The upper limit of Pearson coefficient $\rho=1$ means two similarity matrixes are totally coherent and the corresponding similarity measurement is totally stable. Thus, higher Pearson coefficient $\rho$ would be better.

With each similarity measurement, we calculate the similarity for two data samples with different data amount parameters $\eta$. The results of average bias $\mu$, standard deviation of bias $\sigma$ and the Pearson coefficient $\rho$ of MovieLens, Amazon, Last.FM and Epinions datasets are reported in Fig. 2 (Results of Netflix and Del.icio.us datasets can be found in Supplementary Information). One can easily find that, the PA index is the most stable measurement regardless of the data amount $\eta$, and even with little data, the PA index could give stable evaluation of the object similarity. As the data amount increases, the average biases $\mu$ and the standard deviations of bias $\sigma$ generally decrease. It can be observed that for both the average bias $\mu$ and deviation $\sigma$, the CN, AA and RA indexes have similar decay pattern. When the data amount is little ($\eta<0.1$) the average bias $\mu$ of the CN, AA, RA indexes are almost the highest, and with the increase of $\eta$, the average bias $\mu$ rapidly decrease which means they are sensitive to the data amount. Another dynamic cluster consisting of the COS, SAL, JAC, SOR, HPI, HDI indexes seem to be unsensitive. Although the average bias $\mu$ and deviation $\sigma$ also derease with the increment of data, the decays are much slower than that of the former cluster (the CN, AA, RA indexes). A special measurement refers to the LHN index, which has no apparent dynamic against the data amount $\eta$. Same with the results of average bias $\mu$ and deviation $\sigma$, Pearson coefficient $\rho$ of the PA index is the highest and larger than 0.9 even with the least data amount ($\eta=0.05$). As to the CN, AA, RA indexes, the Pearson coefficients $\rho$ are also sensitive, which is similar to the average bias $\mu$ and deviation $\sigma$. As the data amount $\eta$ increase, Pearson coefficient $\rho$ of the CN, AA, RA indexes rapidly increase to quite high levels. Other measurements' Pearson coefficient $\rho$, however, increases very slowly with the data amount and are in general less than 0.2 even when all the data ($\eta=0.5$) was used. Especially in Amazon which is a very sparse (sparsity is $3.17\times 10^{-5}$) dataset, most measurements' Pearson coefficients $\rho$ are less than 0.03. This result indicates that, for most similarity measurements, the similarity matrixes calculated from different data samples could have no apparent coherence. Overall, more data could make it more stable for most of the measurements especially the CN, AA and RA indexes.

To get deeper insight of the comparison and the classification of those similarity measurements, we analyze the results of average bias $\mu$ and the standard deviation $\sigma$ when all of the data is used ($\eta=0.5$) as shown in Fig. 3 (Results of Netflix and Del.icio.us datasets can be found in Supplementary Information). Using the average bias $\mu$ and the dispersion $\sigma/\mu$ as two dimensions, we can get the $\mu-\sigma$ location map for each dataset respectively. One can surprisingly find that, those similarity measurements could be well classified from the perspective of similarity stability. Except four measurements namely the PA, PC, LHN and IHC indexes, the other measurements could be classified into three clusters. Measurements in the same cluster are similar in both mathematical form and original considerations. The first cluster consists three measurements namely the CN, AA and RA indexes, all of which only take the information of the common neighbors of two target objects into consideration. While the CN index is the total number of common neighbors, the AA and RA indexes make the total number weighted by $1/{\rm log}(k_u)$ and $1/k_u$ respectively where $k_u$ is the degree of the two target objects' common neighbor $u$. The second cluster consists of six measurements namely the COS, SAL, JAC, SOR, HPI and HDI indexes. Except the COS index, the other five measurements are all variations of the CN index, adding the information of the two target objects. However, the CN index's another variation, the LHN index, locates outside the second cluster. The reason may be that, when considering the degree information of two target objects, the LHN index makes the degrees of two objects multiplied, i.e. $k_\alpha k_\beta$, and thus the degree information is quadratic in the LHN index. Unlike the LHN index, other variations's degree information is not quadratic, such as $\sqrt{k_\alpha k_\beta}$ of the SAL index, $k_\alpha + k_\beta$ of the SOR index, $\max(k_\alpha, k_\beta)$ of the HDI index and so on (See Method section for detailed mathematical definitions of these measurements). The third cluster consists the MD and HC indexes which consider the degree information of both the target objects and their common neighbors. Another similar point is that, the MD and HC indexes are both based on the spreading process on bipartite networks according to Physical theories. Actually, they have totally the same stability when measuring the object similarity. Although the basic considerations are different, mathematical definitions of the MD and HC indexes are very similar that leads to $s_{\alpha\beta}^{MD} = s_{\beta\alpha}^{HC}$. Overall, according to the stability of the object similarity, various similarity measurements could be well classified into three clusters. In fact, the classification can also be observed in Fig. 2, in which measurements in the same cluster always have same dynamical pattern against the data amount parameter $\eta$.

\begin{figure}[!htb]
\center\scalebox{0.5}[0.5]{\includegraphics{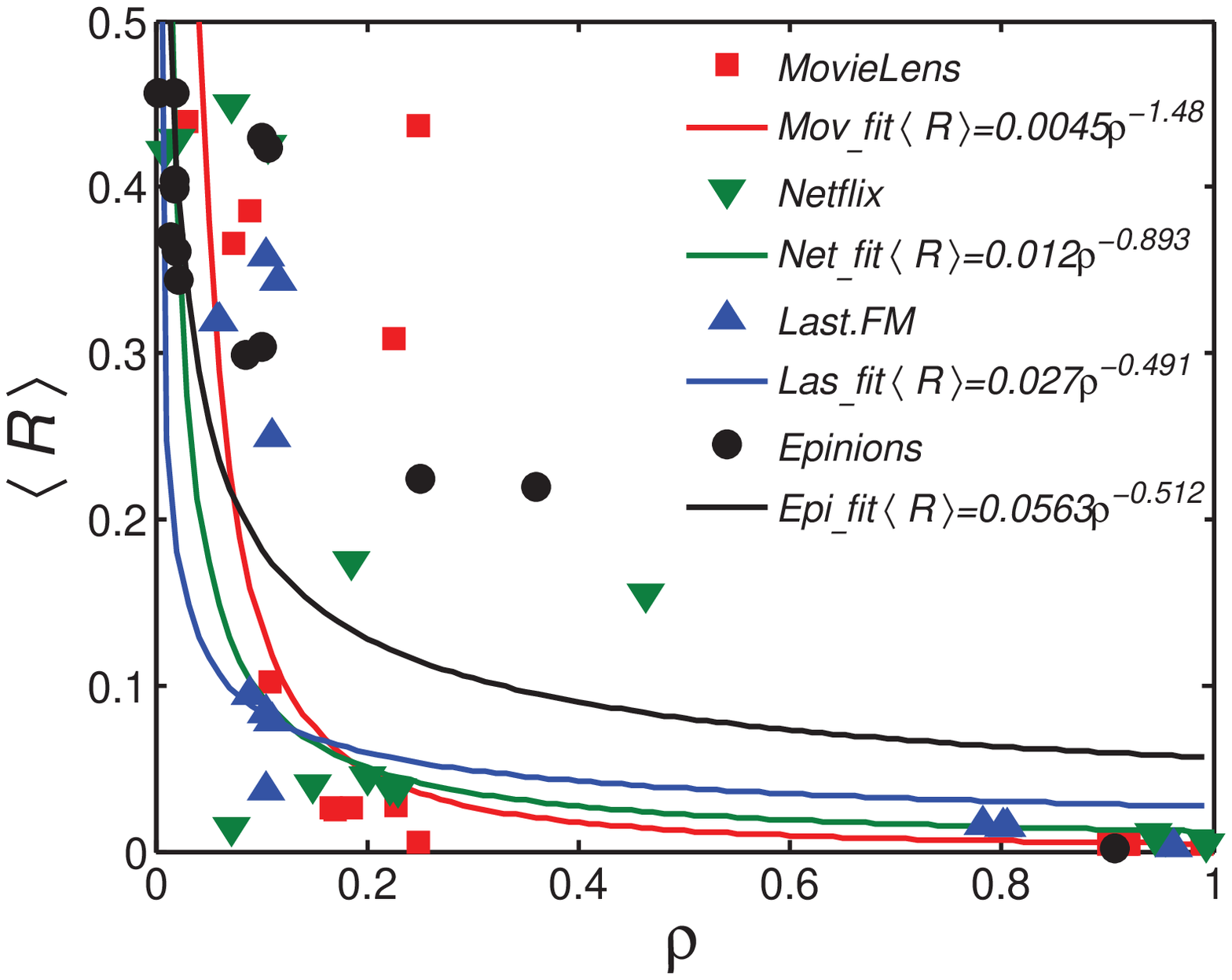}}
\caption{(Color online) The average ranking position $\langle R\rangle$ versus the Pearson coefficient of similarity matrixes $\rho$ for each similarity measurement. The values of $\langle R\rangle$ and $\rho$ are calculated with $\eta=0.5$ and $L=50$ and averaged over 10 independent calculations. Furthermore, we fit the correlation with power-law form $\langle R\rangle=a\rho^{-b}$.}
\end{figure}

\vspace{\baselineskip}
{\large\noindent\textbf{Effect on the Recommendation}}\\
Object similarity in the user-object bipartite networks is generally used to make the recommendation for users. While those fifteen similarity measurements are widely used in the recommendation system, the stability of the recommendation results is still unknown. In this section, we analyze the effect of object similarity stability on the recommendation result. Generally speaking, the goal of a recommendation system is to generate a recommendation list consisted of $L$ objects and voluntarily to display on each user's interface based on the target user's historical selections. The system has to calculate the score of every unselected object for a target user $u$, and rank the objects from high scores to low scores. The score of an object $\beta$ for the target user $u$, $w_{u\beta}$ is given by
\begin{equation}
w_{u\beta}=\sum_{\alpha\in\Gamma_u}s_{\alpha\beta},
\end{equation}
where $\Gamma_u$ is the set of objects which are the historical selections of the user $u$. A high score means that, the system evaluates it as what the target user interests in. Thus, a stable recommendation system should not rank a definite object at totally different positions of the ranking list at different times. To qualify the stability of recommendation results, we still divide the data as two samples according to the former method with $\eta=0.5$. Thus, for a target user $u$, there would be two ranking lists of objects. If an object $\alpha$ is in the recommendation lists (ranking at the first $L$ positions in the ranking list), we define $R_{u\alpha}=i/M$ where $i$ is object $\alpha$'s position in another ranking list and $M$ is the total number of objects. Hence, we can use the average ranking position $\langle R\rangle$ to describe the stability of the recommendation results and $\langle R\rangle$ reads
\begin{equation}
\langle R\rangle = \sum_{u}\sum_{\alpha\in O_u}\frac{R_{u\alpha}}{|O_u|},
\end{equation}
where $O_u$ is the set of objects which rank at the top $L$ positions of the ranking list and at the same time have not been selected by the target user $u$ in both of the samples, and $|O_u|$ is the number of objects in the set $O_u$. According to this definition, stable measurements would have small average ranking position $\langle R\rangle$.

\begin{figure*}[!tb]
\center\scalebox{.92}[.92]{\includegraphics{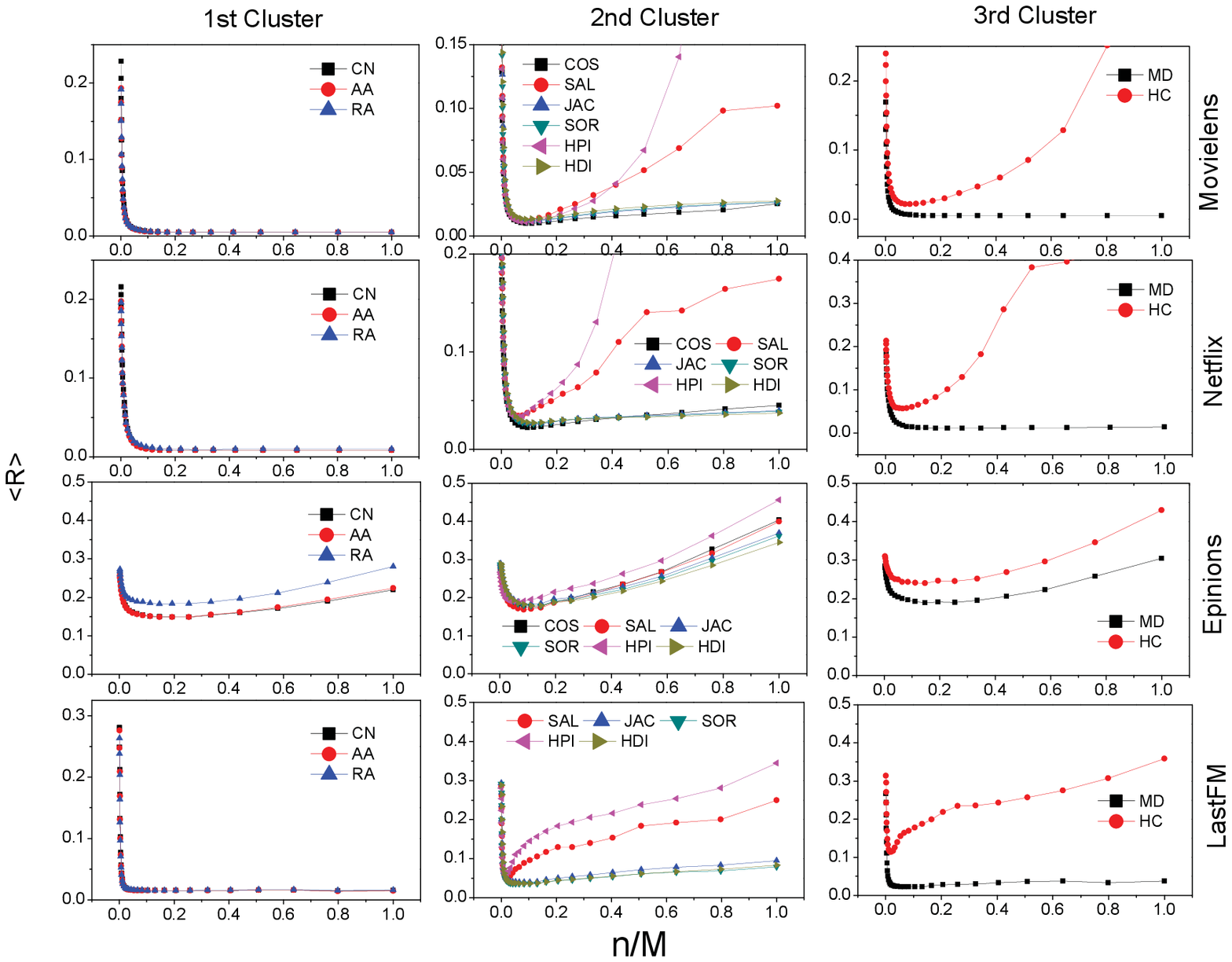}}
\caption{(Color online) The average ranking position of the recommended objects $\langle R\rangle$, against number of objects that counted in the top-$n$-stability method. The length of the recommendation list in the simulation is $L=50$, and the results are averaged over 10 independent simulations. In general, the recommendation stability could be improved by considering only the stable similarities.}
\end{figure*}

To explore the correlation between similarity stability and recommendation stability, figure. 4 shows the Pearson coefficient of similarity matrixes $\rho$ and the average ranking position $\langle R\rangle$ of each measurement for MovieLens, Netflix, Last.FM and Epinions datasets. We can observe the correlation that, the more stable a measurement could evaluate the similarity (higher $\rho$), the more stable its' recommendation would be (lower $\langle R\rangle$). However, the correlation is not linear. We fit the results of each dataset with power-form $\langle R\rangle=a\rho^{-b}$. Although the parameters in the fitting equation may be different, it could be concluded that, there is a power-law correlation between similarity stability and recommendation stability. The difference of the parameters may be caused by the different structures and sparsity of the data.

On the other hand, we can find that, many of the recommendations are quite unstable, such as the PC, SAL, HPI, LHN, HC and IHC indexes whose average ranking positions $\langle R\rangle$ are larger than 0.1 in each dataset (See Table S1 for detail). To take MovieLens dataset as an example, $\langle R\rangle=0.1$ means that, the objects recommended at the top $L$ positions using a data sample are ranked at 585th position (there are 5850 objects in MovieLens data) at average when using another data sample. Theoretically, the average ranking position $\langle R\rangle$ of the totally random case is 0.5 and the most stable results is $\langle R\rangle=L/2M$ where $M$ is the total number of the objects. Thus, the theoretical best stability is $4.3\times10^{-3}$, $4.9\times10^{-3}$, $8.3\times10^{-4}$ and $1.4\times10^{-3}$ for MovieLens, Netflix, Epinions and Last.FM respectively which means the recommendation lists of the two data samples are totally the same.

To improve the stability of the recommendation and further explore the effect of the similarity stability, here we propose a top-$n$-stability method. For an object $\alpha$, the similarity bias between $\alpha$ and each object $\beta$, $\delta_{\beta\alpha}$ is calculated and ranked from low (stable) to high (unstable). When adding the score of object $\alpha$ according to Eq. (4), we only take $n$ objects which have the most stable similarities i. e., which rank at the top $n$ positions to object $\alpha$. This could be explained as
\begin{equation}
s_{\beta\alpha}=\left\{
\begin{array}{l}
s_{\beta\alpha}~~~~~~~ P_{\beta\alpha}\leq n \\
0 ~~~~~~~~~~~ P_{\beta\alpha}>n
\end{array}
\right. ,
\end{equation}
where $P_{\beta\alpha}$ is the position of object $\beta$ in object $\alpha$'s stability list. Note that, unlike the classical top-$n$-similarity recommendation algorithm in which $n$ objects with the highest similarities to object $\alpha$ would be counted \cite{TOPN1, TOPN2}, here we ignore the exact value of similarity, just consider the stability. The basic consideration is that, if a pair of objects' similarity has poor stability, the similarity would be meaningless regardless of the value of similarity. Through the experiments, the classical top-$n$-similarity method can also improve the recommendation's stability for a little bit, but the improvement of our top-$n$-stability method is much bigger (See Supplementary Information).

With different number of stable objects $n$, we make the recommendation and calculate the average ranking position of the recommended objects $\langle R\rangle$ as shown in Fig. 5. Figure 5 is summarised according to the similarity measurements' clusters, the results of similarity measurements the PC, LHN, PA and IHC indexes could be found in Supplementary Information. One can observe that, there is no apparent recommendation stability improvement for the first cluster (the CN, AA, RA indexes) except in the Epinions dataset in which the recommendation stability is poor for every similarity measurement. On the other hand, recommendation stability of measurements of the second cluster could be well improved by the top-$n$-stability method especially for the SAL and HPI indexes whose average ranking position $\langle R\rangle$ is over 0.1. However, measurements in the third cluster, i.e. the MD and HC indexes, have different pattern against the top-$n$-stability method. While the HC index's recommendation stability could be largely improved, the MD index has no apparent improvement. We can conclude that, when the recommendation is unstable, our top-$n$-stability method could largely improve (See Table S2 for detailed improvement ratio) the stability by taking only the stable similarities into account. For most similarity measurements, the optimized stability could be reached when considering about 10\% of the similarities, and for the poor-stability measurements, the counted ratio may even be about 5\%. The improvement indicates that, those unstable similarities are more like false information which would lead to the deflected evaluation of users' true interest.

\vspace{\baselineskip}
{\large\noindent\textbf{Conclusion and Discussion}}\\
While similarity measurements can measure the potential relations between objects in the biological, social, commerce systems, they are meaningful only if the evaluated similarities are stable. Unstable similarities are generally false information which would lead to the misunderstanding of the relations between objects. The present paper studied the stability of fifteen similarity measurements measuring object similarity in user-object bipartite networks. The results showed that, most similarity measurements except the PA, CN, AA, RA indexes, are quite unstable when measuring object similarity. The Pearson coefficient $\rho$ of two similarity matrixes calculated from two data samples may even smaller than 0.2, which means the two matrixes have little correlation. Generally speaking, measurements with simple considerations can describe the natural properties of objects and are stable. The CN, AA, RA indexes considering only the information of two objects' common neighbors are stable and can be regarded as a cluster. On the other hand, variations of the CN index, namely the SAL, JAC, SOR, HPI, HDI indexes, considering further the degree information of two objects, are less stable than the CN index and can be regarded as another cluster. Measurements in the same cluster have in general similar considerations and mathematical definitions, and thus have similar stabilities and even the dynamic against the data amount. In other words, while dozens of measurements have been developed, those similarity measurements can be well classified according to their object similarity stability. When a new measurement being proposed, one just needs to analyze its stability to identify which cluster it belongs to, and then could get deeper insight to this measurement by comparing with other measurements within the same cluster. In addition, we presented a top-$n$-stability method to investigate the effect of object similarity on the recommendations. By considering only the stable similarities i.e. deleting the unstable, false information, the stability of the recommendation could be improved.

The investigations and considerations in this paper only focused on the objects. Actually, similarity is also an important method measuring the potential relations of human beings in the social systems and users in the online systems \cite{POP_SIM}. However, different with objects whose natural properties are definitely unchangeable, evidences have been found to prove that, the behaviors and interests of human behavior are temporal \cite{TEMPROAL1, TEMPROAL2}. Thus, the stabilities of object similarity and human-to-human similarity may have totally different meanings. Additionally, the stability of those similarity measurements should be also studied in one-mode systems, which contain only one kind of nodes. Especially for the objects like genes, proteins ect., the investigations of similarity stability are still urgently needed because those objects may have properties different from that studied in this paper.
\\ \\
\vspace{\baselineskip}
{\large\noindent\textbf{Methods}}\\
The datasets used in the present paper are usually modeled as user-object bipartite network in which nodes can be divided into two groups, representing users and objects respectively. In such kind of system, links only exist between different kinds of nods, i.e. between users and objects. We use $\alpha$ and $\beta$ to denote the target pair of objects and $U_{\alpha\beta}$ is the set of users who select both objects $\alpha$ and $\beta$. The popularity $k_\alpha$ and $k_\beta$ represent the times of object $\alpha$ and $\beta$ selected by users respectively, and the activity $k_u$ is the number of objects user $u$ have selected. We suppose that, the function ${\bf min}(x, y)$ equals to the minimum value between $x$ and $y$ and ${\bf max}(x, y)$ equals to the maximum value between $x$ and $y$. In addition, ${\bf r_\alpha}$ and ${\bf r_\beta}$ are rating vectors in the N-dimensional user space and $r_{u\alpha}$ and $r_{u\beta}$ is the rating user $u$ evaluating the object $\alpha$ and $\beta$ respectively. With those parameters defined, the fifteen similarity measurements referred in this paper read:
\begin{align*}
COS: & ~~~ s_{\alpha\beta}=\frac{1}{|{\bf r_\alpha}||{\bf r_\beta}|} \sum_{u\in U_{\alpha\beta}}r_{u\alpha}r_{u\beta}, \\
PC: & ~~~ s_{\alpha\beta}=\frac{\sum_{u\in U_{\alpha\beta}}(r_{u\alpha}-\overline{r}_\alpha)(r_{u\beta}-\overline{r}_\beta)}{\sqrt{\sum_{u\in U_{\alpha\beta}}(r_{u\alpha}-\overline{r}_\alpha)^2} \sqrt{\sum_{u\in U_{\alpha\beta}}(r_{u\beta}-\overline{r}_\beta)^2}},\\
CN: & ~~~ s_{\alpha\beta}=\sum_{u\in U_{\alpha\beta}}1,\\
SAL: & ~~~ s_{\alpha\beta}=\frac{1}{\sqrt{k_\alpha k_\beta}}\sum_{u\in U_{\alpha\beta}}1, \\
JAC: & ~~~ s_{\alpha\beta}=\frac{1}{k_\alpha+k_\beta-\sum_{u\in U_{\alpha\beta}}1}\sum_{u\in U_{\alpha\beta}}1,\\
SOR: & ~~~ s_{\alpha\beta}=\frac{2}{k_\alpha+k_\beta}\sum_{u\in U_{\alpha\beta}}1,\\
HPI: & ~~~ s_{\alpha\beta}=\frac{1}{{\bf min}(k_\alpha, k_\beta)}\sum_{u\in U_{\alpha\beta}}1,\\
HDI: & ~~~ s_{\alpha\beta}=\frac{1}{{\bf max}(k_\alpha, k_\beta)}\sum_{u\in U_{\alpha\beta}}1,\\
LHN: & ~~~ s_{\alpha\beta}=\frac{1}{k_\alpha k_\beta}\sum_{u\in U_{\alpha\beta}}1,\\
AA: & ~~~ s_{\alpha\beta}=\sum_{u\in U_{\alpha\beta}}\frac{1}{\log k_u}, \\
RA: & ~~~ s_{\alpha\beta}=\sum_{u\in U_{\alpha\beta}}\frac{1}{k_u}, \\
PA: & ~~~ s_{\alpha\beta}=k_\alpha k_\beta,\\
MD: & ~~~ s_{\alpha\beta}=\frac{1}{k_\alpha}\sum_{u\in U_{\alpha\beta}}\frac{1}{k_u}, \\
HC: & ~~~ s_{\alpha\beta}=\frac{1}{k_\beta}\sum_{u\in U_{\alpha\beta}}\frac{1}{k_u}, \\
IHC: & ~~~ s_{\alpha\beta}=\frac{1}{k_\beta^2}\sum_{u\in U_{\alpha\beta}}\frac{1}{k_u}.
\end{align*}

\begin{addendum}

\item  This work is partially supported by NSFC (71371125, 61374177 and 71171136), Shanghai Leading Academic Discipline Project (Systems Science) (No. XTKX2012), JGL is supported by the Program for Professor of Special Appointment (Eastern Scholar) at Shanghai Institutions of Higher Learning.

\item[Author Contributions] L.H. and J.G.L. conceived the idea. L.H. and X.P. performed the calculation. L.H., X.P., Q.G., J.G.L. and T.Z discussed the results and wrote the manuscript. Correspondence and requests for materials should be addressed to J.G.L. (liujg004@ustc.edu.cn).

\item[Competing Interests] The authors declare that they have no competing financial interests.

\end{addendum}
\ifthenelse{\boolean{SubmittedVersion}}{\processdelayedfloats}{%
\cleardoublepage}

\end{document}